# Structural network heterogeneities and network dynamics: a possible dynamical mechanism for hippocampal memory reactivation

Authors: Piotr Jablonski[1,2], Gina R. Poe[3], Michal Zochowski[1]

1. Department of Physics and Biophysics Research Division
   University of Michigan, Ann Arbor, MI 48109

2. Institute for Social Studies
   University of Warsaw, Warsaw, Poland

3. Department of Anesthesiology and Department of Molecular and Integrative Physiology
   University of Michigan Medical School, Ann Arbor, MI 48109-0615

## Abstract

The hippocampus has the capacity for reactivating recently acquired memories [1-3] and it is hypothesized that one of the functions of sleep reactivation is the facilitation of consolidation of novel memory traces [4-11]. The dynamic and network processes underlying such a reactivation remain, however, unknown. We show that such a reactivation characterized by local, self-sustained activity of a network region may be an inherent property of the recurrent excitatory-inhibitory network with a heterogeneous structure. The entry into the reactivation phase is mediated through a physiologically feasible regulation of global excitability and external input sources, while the reactivated component of the network is formed through induced network heterogeneities during learning. We show that structural changes needed for robust reactivation of a given network region are well within known physiological parameters [12,13].

## 1. Introduction

It has been shown that the hippocampus is able to generate experience-dependent reactivation during REM sleep [1-4,8,9], non-REM [1,2,5,6,14-16] sleep and during quite periods of waking [3]. The spatio-temporal patterning of neuronal activity during reactivation is correlated with the patterning of the preceding wake activity [4] and the correlation between cells coactive during waking is also higher during sleep [5,15,17,18]. It is widely thought that this reactivation serves as a mechanism of memory consolidation [8,19,20]. The specific reactivation mechanisms mediating memory storage remain unknown, though the unique neurochemical environment [8] and gene expression patterns [21,22] during REM favors a bidirectional synaptic plasticity role for this state [8].

In this paper we address the question of what dynamical mechanisms are associated with memory reactivation and, furthermore, what network features mediate the formation of robust reactivation of network regions. We show that such a reactivation characterized by local, self-sustained activity of a network region may be an inherent property of the recurrent excitatory-inhibitory network with a heterogeneous structure.

The relationship between network topology and dynamics has been recently extensively studied for social, biological and artificial networks, including the synchronizability and coherence of elements in the network [23-29]. It has been shown that synchronization becomes optimal in a system exhibiting Small World (SW) topology [30-32]. However the introduction of heterogeneity in the nodal degrees reduces the ability of the network to achieve global synchronization. At the same time it was shown that local variations of network connectivity characterized by differences in nodal degrees or coupling strength produce a hierarchical synchronization that can be observed on separate timescales [33,34].

It has also been shown that a single layered network of excitatory neurons, having SW topology generates bistable behavior between self-sustained activity and the quiescent state [23].

We show that the heterogeneity of the excitatory network together with the interplay between locally propagating, recurrent excitation and global inhibition provide a medium to obtain selective regions of self-sustained activity in the network. The regions of the network having a higher degree or stronger connectivity locally change intrinsic excitability of this region and have a lower threshold of transition to self-sustained recurrent activity. Thus the location of reactivating regions in the network is inherently determined by the location of those fluctuations.

We illustrate these dynamics on a simple SW network that structurally emulates CA hippocampal network consisting of two interacting layers composed of excitatory and inhibitory neurons, which represent the local networks of pyramidal cells and interneurons, respectively. In both types of local networks short local connections coexist with shortcuts between distant neurons. However the topology of the excitatory, pyramidal cell network is skewed towards local connectivity, while the connectivity between inhibitory interneurons is composed predominantly of random shortcuts.

To test the validity of the obtained network dynamics we made recordings of tens of simultaneously recorded CA1 cells in the freely behaving rat [17] in order to compare the activity patterns during waking exploration and during spontaneous reactivation in sleep with those of our model network. Rats implanted weeks before with a headstage multiple single unit recording system [18] were run on a novel maze for food reward while hippocampal CA1 place cell activity was gathered. The same cells were again recorded for more than 4 hours of subsequent spontaneous sleep. The hippocampus generated experience-dependent reactivation as observed by others during REM sleep [1-4,8,9], non-REM sleep [1,2,5,6,14,15,21] and periods of quiet waking [3]. The spatio-temporal patterning of neuronal activity during reactivation is correlated with the patterning of the preceding learning activity [4].. In the present study, firing rates and patterns of rat hippocampal neurons were measured across sleep/waking reactivation states and compared with model network dynamics. For a detailed experimental protocol please refer to [9,17].

**2. Network structure and dynamics**

The network is composed a larger population ($N_e = 500$) of excitatory neurons and smaller population ($N_i = 100$) of inhibitory neurons. This specific cell ratio is chosen to emulate the physiology of the CA network but does not affect the observed phenomenon. Both inhibitory and excitatory networks are one-dimensional SWN having periodic boundary conditions. For the excitatory network, the local connections are established within $R_e = 10$ radius and $p_g^e$ is the rewiring parameter. Similarly, the interneuron subpopulation has $R_i = 2$ and $p_g^i = 1$, forming a random graph network with global connections. Every inhibitory interneuron receives input form $n_{ei} = 5$ neighboring excitatory neurons, while every excitatory neuron receives input from $n_{ie} = 10$ randomly chosen inhibitory ones. Thus the functional topology of the whole system provides global inhibition driven by focal excitation in the network.

The membrane potential for individual neurons is determined by:

$$C\frac{dV_{i/e}^j}{dt} = -\alpha_j V_{i/e}^j + I_{i/e} + \sum_k w^{jk} I_{syn}^k \quad (1)$$

where subscript $i/e$ denotes the type of the neuron (inhibitory or excitatory, respectively) and $V_j$ is the membrane voltage of the j-th cell. When $V_j \geq 1$ the neuron fires an action potential and the membrane potential is reset to $V_j = 0$. The $\alpha_j \in [1,1.3]$ is the membrane leak time constant, $C$ is membrane capacitance. The $I_{i/e}$ parameter controls global excitability for excitatory and inhibitory networks. The $w^{ij}$ term denotes the strength of the synaptic connections from i-th to the j-th neuron. The strengths are constant within and between cell populations, so that: $w_{e/e} = 2.0$, $w_{i/i} = 10.0$, $w_{e/i} = 4.0$, and $w_{i/e} = 2.0$, where subscript $i/e$ denotes the type of originating/target population, respectively. These specific values were chosen to balance amount of excitation and inhibition in the network, but otherwise do not influence observed dynamics. Finally, $I_{syn}^i$ is a postsynaptic current adopted from Netoff [27] and given by:

$$I_{syn}^i(t) = \exp\left(\frac{-(t - t_{spike}^i)}{\tau_f}\right) - \exp\left(\frac{-(t - t_{spike}^i)}{\tau_s}\right) \quad (2)$$

where $t_{spike}^i$ is the time of spike of the i-th neuron and $\tau_f = 3$ and $\tau_s = 0.3$. The same type of network behavior can be obtained using pulse coupling [23].

In addition to the dynamics described by Eqn.1, independent of its membrane potential every neuron may fire an action potential at every time step with a probability ($p_{external} = 0.0003$ if not stated otherwise). This process is to mimic neuronal activity that is caused by activity independent of the network dynamics.

Local network heterogeneity is obtained by the addition of connections (i.e. local changes in nodal degree) or by strengthening the already existing ones between random groups of neurons. Those connectivity fluctuations are in principle caused by synaptic modifications incurred during the learning process [35,36]. For visualization purposes we choose five independent regions, each constituting a single "memory subnetwork".

The network dynamics fall into three basic regimes (Fig 1), with the global excitability of inhibitory ($I_i$) and excitatory ($I_e$) networks being the control parameters of the transitions between these regimes (Fig. 1B):

**Random activation** (Fig 1B, region 1, Fig 1C and 1F) takes place for low global excitability of the network ($I_e < I_i$ and $I_e, I_i < 0.7$). In this regime all neurons fire infrequently and do not generate the activation cascades needed to obtained self-sustained activity. The network activity is homogeneous and therefore independent of structural heterogeneities. This is due to the fact that the neurons' membrane potential is too far from the threshold.

**The local self-reactivation regime** (Fig. 1B, region 2, Fig 1D and 1G) occurs at intermediate network excitability levels. In this regime, regions of the network can spontaneously reactivate (Fig 2), generating self-sustained cascades of neural activity [23]. This persistent activity relies on a reinjection of activity through existing shortcuts into an already recovered, previously active domain. This is due to the fact, that the network regions containing additional connectivity have higher intrinsic excitability, allowing local spontaneous reactivation for lower values of global network excitability. Figure 2 depicts the occurrence of reactivation dynamics as a function of network excitability, for network regions that have various numbers of supplementary connections added. The reactivation threshold in terms of global network excitation ($I_e, I_i$) is a decreasing monotonic function of the number of supplementary connections added.

The initial activation of a given network region (memory) happens at random when enough neurons co-activate to generate the cascade of local network activity. This activation is characterized by a significantly higher total firing rate and, in turn, increased activity of inhibitory neurons. The increased level of inhibition spreads throughout the excitatory network through the inhibitory feedback connections, limiting other regions in the network from simultaneously reactivating (Fig 2B-D). However, due to spontaneous activity fluctuations within the other network regions and/or a finite probability of internal failure [23] of the already active region, there is a finite probability that the already reactivated region can be replaced by reactivation of another one. These processes result in time-specific local activation such that the overlap between the activated regions is small (Fig. 1G).

Thus, the interplay of two basic mechanisms is driving time-resolved reactivation: local fluctuations in network excitability that allow for focal reactivation, and the interaction of local network excitability and global inhibition that is limiting reactivation of other network regions while one region is reactivated. The last mechanism is similar in its basic function to the lateral inhibition observed in different brain modalities [37].

**The bursting regime** (Fig. 1B, region 3, Fig. 1E end 1H) occurs when excitability in the network is high ($I_e > I_i$) and leads to global coactivation of excitatory cells, independent of the network region to which they belong, resulting in highly periodic synchronous bursts of activity across the network.

**3. Magnitude of network heterogeneities leading to local self-sustained activation.**

We investigated the magnitude of local network heterogeneities that must occur in a given region of the network in order to facilitate robust patterns of reactivation. Such structural modifications are thought to take place during the initial memory formation mediated by the potentiation of existing synapses and/or the formation of new ones.

We initially created a network having no embedded memories, i.e., the topology and connectivity strengths were uniform (Fig. 3B). To facilitate the activation of neurons from a given region constituting a memory, we added connections (Fig. 3A) or strengthened the coupling between existing ones (Fig 3E). We then measured the degree of activation of the network region as a function of the added number or increased strength of connections: we calculated what fraction of activity (number of spikes within the time window) of the whole network is generated by the neurons constituting that memory. A high fraction of activation (70-90%) of any one memory region indicates robust reactivation of that memory, whereas a fraction of ~ 20% indicates a lack of reactivation.

The number of additional connections or the magnitude of increase in the coupling strength depends strongly on the topology of the underlying global network. For global network topologies favoring dense local connectivities, a relatively small modification of the network region (i.e., low number of additional connections added or a small increase of coupling strength) produces robust memory reactivation. For topologies favoring global connectivities, larger local modifications are needed to generate a region of robust reactivation. Figures 3C&D show the reactivation resulting from an increased number of connections within the memory and from a larger coupling strength, respectively. Both, the doubling in coupling strength as well as a 2% increase of local connectivity (measured with respect to a fully connected network) within the memory subnetwork are well within neurophysiological limits [12,13].

**4. Effects of random input on the self-activation of neuronal ensembles.**

We investigated the changes in network dynamics in response to changing levels of external inputs (EI) to the network, which was obtained by varying the probability of neurons firing ($p_{external}$) independently of the internal network state. Higher levels of $p_{external}$ result in a higher mean activation in the whole network. However, the increased $p_{external}$ does not induce self-sustained activity in the network, but abolishes it (Fig. 4). Thus, the internal network reactivation is effectively inhibited by external inputs. This is due to the fact that the random activity of the network does not allow for formation of propagating active domains that underlie the self-sustained activity.

To quantify this effect we calculated the mean spike frequency of all neurons in the network and its standard deviation (SD) (Fig. 4A). When network dynamics facilitate memory reactivation (low-EI), the mean spike frequency is relatively lower but its SD is large (Fig. 4A) because of the coexistence of two distinct populations in the network: neurons belonging to the reactivated region, with a much higher spiking frequency, and the remaining neurons with lower spiking frequency (Fig. 4Ba , 4C). For higher EI levels, the memories are progressively less activated, resulting in a uniform Gaussian distribution of firing frequencies (Fig 4Bb,c, 4D).

This finding is consistent with known experimental measurements. It has been shown that during sleep there is inhibition at the thalamus of the neuronal pathways leading from various sensory modalities into the neocortex and hippocampus [38-41].

We have also compared the firing rate distributions obtained from our model with the ones obtained experimentally from *in vivo* recordings of foraging and sleeping rat hippocampal cells. Figure 4E presents the experimentally obtained distributions of instantaneous firing frequencies of CA1 pyramidal cells when rats freely explore a novel environment (top; see supplemental material) and during subsequent REM sleep (bottom) when this newly acquired experience was reactivated. The model network reproduces the experimentally observed shift toward higher firing frequencies during waking, when thalamic sensory inputs were high. The model also reflects the shape of the rat CA1 cell firing frequency distribution both in REM sleep and waking.

**5. Activation of a single memory region through the application of external bias.**

Initial memory formation and subsequent memory reactivation require radically different dynamics. During memory formation, the externally stimulated neural pattern has to be incorporated into the network dynamics. Activation of additional neurons not directly associated with that memory would be undesirable as it would result in spurious correlations that would impair the quality of the newly formed memory. On the other hand, during memory reactivation or retrieval, the activation of relatively few neurons should lead to the reactivation of many others also involved in the associated memory.

We investigated how the application of a bias to the given memory would facilitate activation of that memory under different network excitability (Fig. 5A) and external input values (Fig. 5B). The bias was introduced by increasing the excitatory drive $I_e$ to a varying number of randomly chosen neurons constituting the reactivating region. As before, to quantify the degree of activation of the region, we calculated what fraction of activity (number of spikes within time window) of the whole network is generated by the neurons constituting that memory region.

For low values of neuronal excitability, the application of varying bias does not change activation within the region, and the curve remains flat. With an increased level of excitability, the biased memory region becomes progressively more excited, forming a sigmoid activation function. For high excitability levels (e.g., REM), the bias applied even to a few neurons immediately activates the whole region (Fig. 5A).

Conversely, in Figure 5B, increasing external inputs ($p_{external}$) made it progressively harder to obtain local activation. For low values of $p_{external}$, the memory region activates for low values of bias. However, for large values of $p_{external}$, the progressive bias only weakly increases regional activation.

**6. Summary and discussion**

We have shown that memory reactivation dynamics observed in the hippocampus during sleep could be an inherent property of the interacting inhibitory and excitatory networks characterized by the coexistence of local connectivity and random shortcuts. We show that network heterogeneities obtained through the creation of new synapses or the strengthening of existing ones, lead to robust self-sustained activation of the neurons constituting the restructured region of the network (memory). The entry into the reactivation phase is mediated through physiologically feasible regulation of the global excitability of the network and the strength of the external input sources. The time resolved reactivation of subsequent regions is mediated through the interplay of local excitation and global inhibition in the network.

From neurophysiologic perspective, since external input can inhibit internal reactivation, e.g. during the waking phase, we hypothesize that sleep provides the conditions for successful reactivation in two ways: sleep can increase network excitability through an increase in acetylcholine levels (i.e., during REM) and at the same time sleep limits external inputs to the network. The reactivation patterns shown in the simulations were corroborated by experimental data gathered from learning and sleeping rats.

**Acknowledgements** This work was supported by NIH EB003583 (MZ), NIH MH60670 (GP) and by a Fulbright fellowship (PJ).

**Author Information** Correspondence and requests for material should be addressed to M.Z. ([michalz@umich.edu](mailto:michalz@umich.edu)).

**Figure Captions**

**FIGURE 1.** Three dynamical regimes can be observed in network behavior. A) Diagram of the investigated network. B) Overlap of activated network regions as a function of excitability of inhibitory ($I_i$) and excitatory ($I_e$) network. The network achieves three distinct dynamical regimes: 1. Cells fire randomly, driven by external sources; 2. Local self-reactivation regime; 3. The synchronous bursting regime. The numbers 1, 2, and 3 denote parameter regimes that are later used in panels C,D,E,F,G,H.

C) Raster of firing pattern of excitatory neurons within regime 1; D) regime 2, E) regime 3. Black dots depict the time of action potentials.

F) Total activation of the excitatory network calculated as $A(t) = \sum_i I_{syn}^i(t)$ within regime 1;

G) regime 2; H) regime 3.

Mean overlap of the local activations, was calculated based on estimation of instantaneous frequencies of observed neuronal activity within each region. We calculated the spike frequency within M consecutive time windows for every network region; M constitutes dimension of the formed activation vector. The activation vectors were normalized and dot product of each pair was calculated.

**FIGURE 2.** Dynamics of reactivation of a single network region: A) Occurrence of reactivation dynamics (the state when a single memory generates over 50% of the total spikes) as a function network excitability for network regions with varying numbers of supplementary connections. B) Raster plot of neural activity during the reactivation dynamics. C) Fluctuations in total excitatory current due to local reactivation. D) Global feedback inhibition of the excitatory network during reactivation. Local fluctuations in **network excitability reduce the reactivation threshold and allow for focal reactivation. The interaction of local network excitability and global feedback inhibition limits the reactivation of other network regions while one region is reactivated.**

**FIGURE 3. Measurement of local structural inhomogeneity required for robust self-activation of network region. A) Fraction of the total network activity observed within a formed memory as a function of added connections for different topologies of the excitatory network. B) Example of the dynamics in the self-activation regime in homogeneous network. C) Example of network dynamics with increased connectivity strength inside a single region. D) Example of the network dynamics with increased connectivity within the memory region. The number of new connections is 5% of that in a fully connected network ($n(n-1)$, where n is the number of neurons constituting a region). E) Fraction of the total network activity observed within a formed memory region as a function of a fractional change in connection strength within that region for different topologies of the excitatory network (as defined above. Error bars show the standard deviation, n=10.**

**FIGURE 4. Activation of the network as a function of external input to the network – simulations and experimental data.**
**A) Average frequency and standard deviation of the mean spiking frequency of all excitatory neurons in the network. The frequency distributions are normalized by dividing them by mean firing frequency obtained for $p_{external} = 0.0003$. External activation of the network (high p$_{external}$) inhibits self-reactivation of any network region even though the network is on average more activated. B) Examples of the distributions of the normalized firing frequencies**

for the three values of external input a, b, and c of (A). The distribution changes from a) bimodal distribution ($p_{external} = 0.0003$), to b) Gaussian distribution for high levels of incoherent activation ($p_{external} = 0.0009$),and c) $p_{external} = 0.003$.

C, D) Examples of the raster plots for low (C, $p_{external} = 0.0003$) and high (D, $p_{external} = 0.003$) level of external input).

E) Experimentally obtained distributions of instantaneous firing rates during waking (bottom) novel maze learning and during novel maze reactivation in REM sleep (top) Frequency is normalized by the mean frequency of REM sleep activity.

FIGURE 5. Activation of the biased network region as a function of network parameters. Local activation is calculated as a fraction of the activity of the reactivating region and the total activation of the excitatory network.

A) Activation as a function of excitability of excitatory and inhibitory networks ($I = I_e = I_i$). For low values of $I$, increasing the fraction of stimulated neurons does not significantly change local activation. For higher values of $I$, the whole region activates. B) Local activation as a function of external input ($p_{external}$) to the network. For higher levels of input, the local response is proportional to the number of stimulated neurons (no activation). For low levels of noise the whole activates, rapidly facilitating local reactivation.

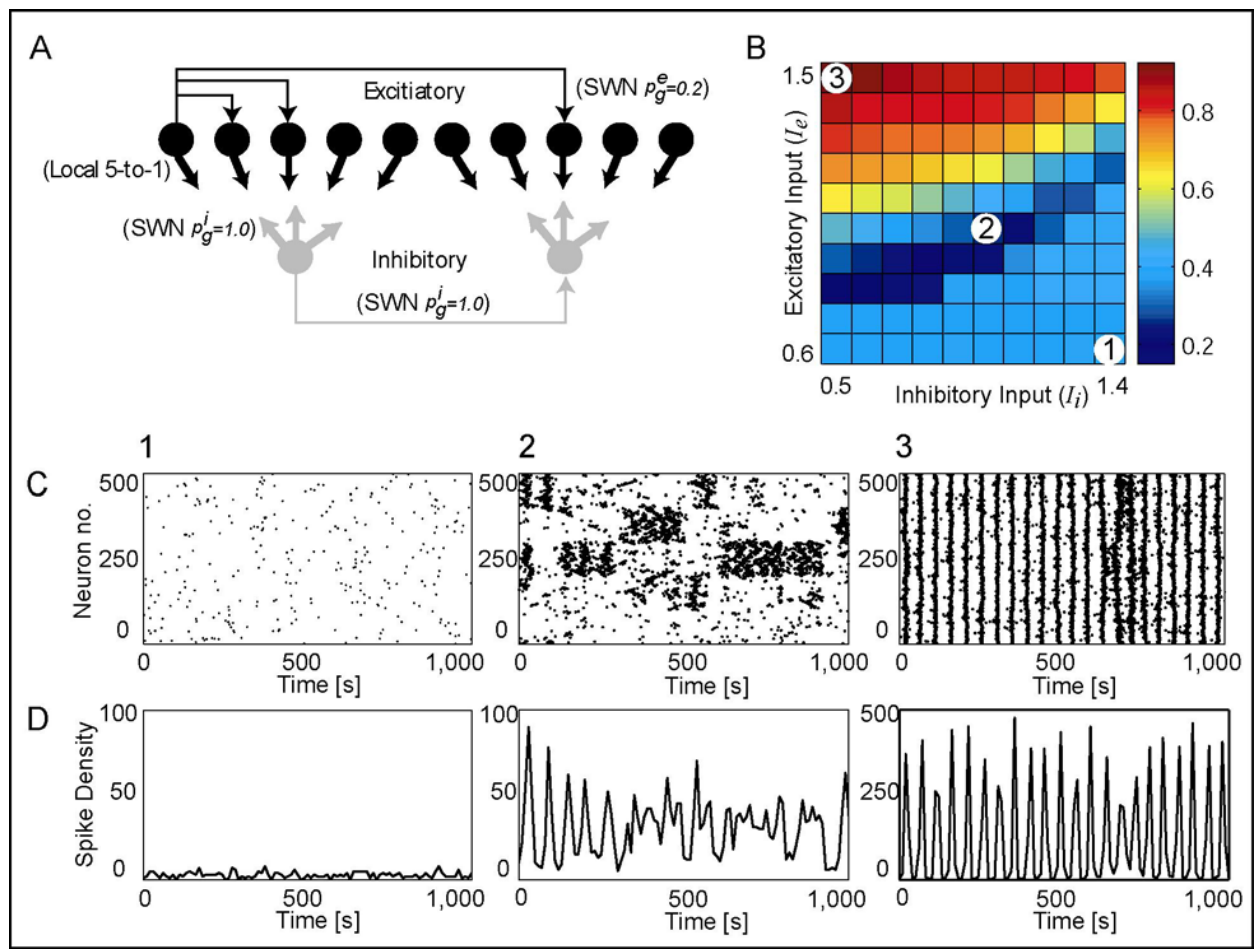

**FIGURE 1**

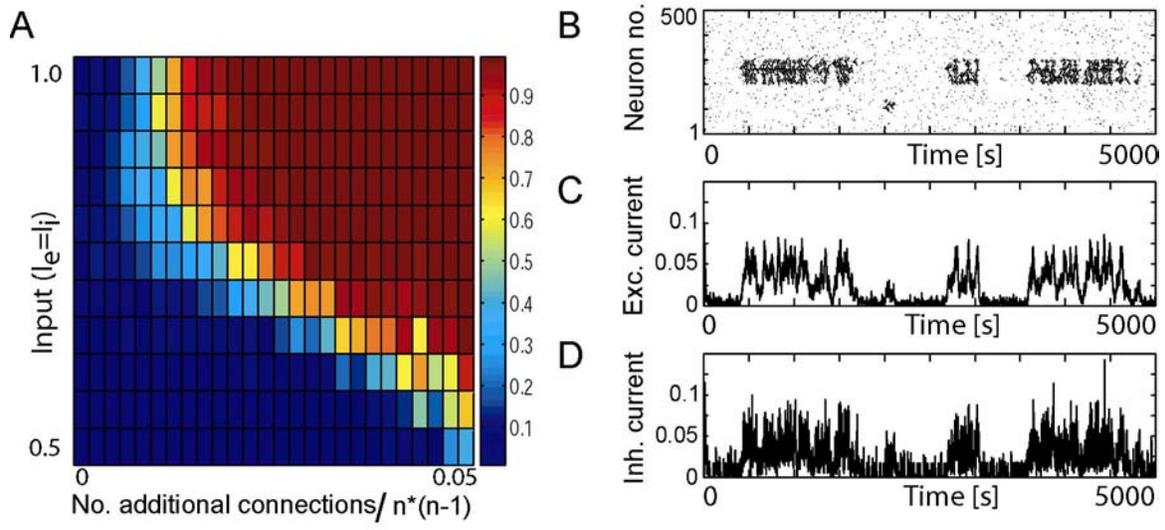

**FIGURE 2**

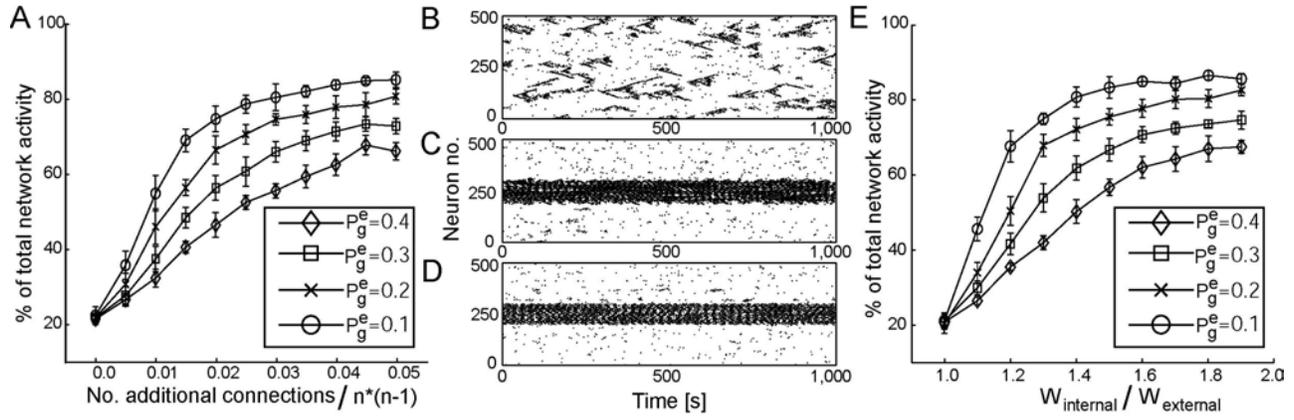

**FIGURE 3**

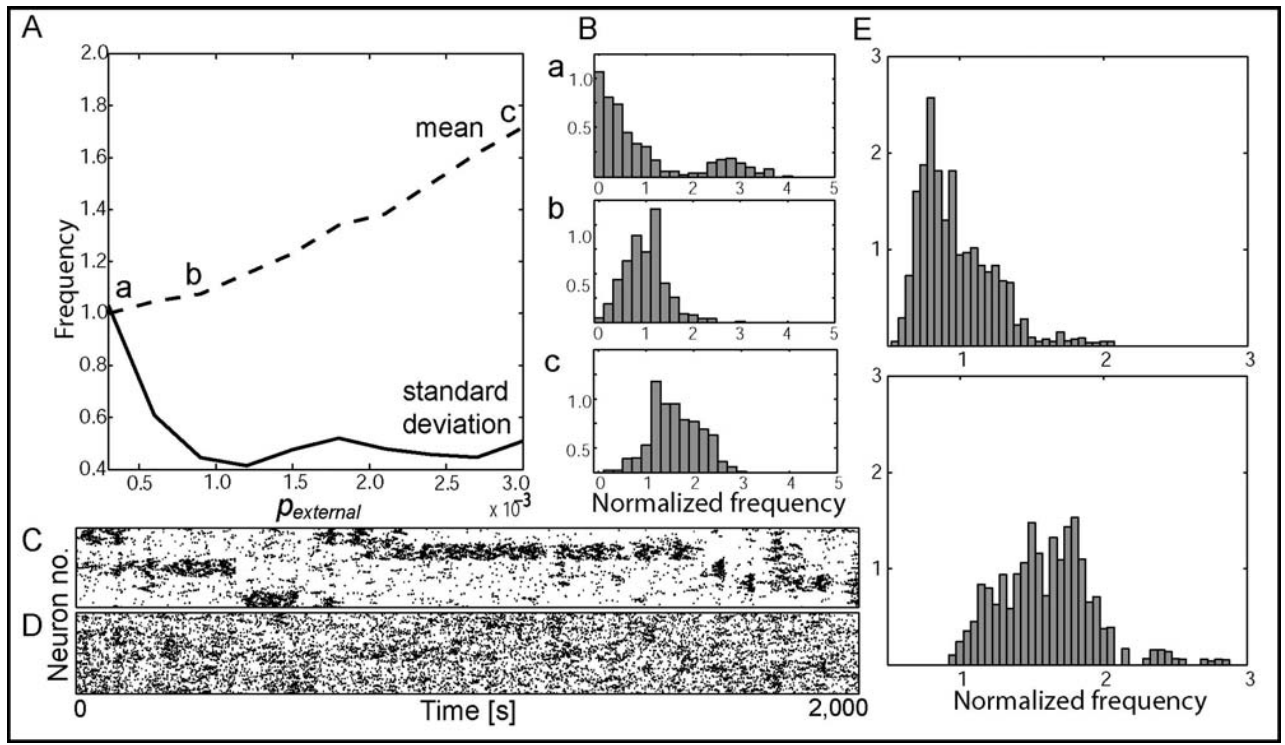

**FIGURE 4**

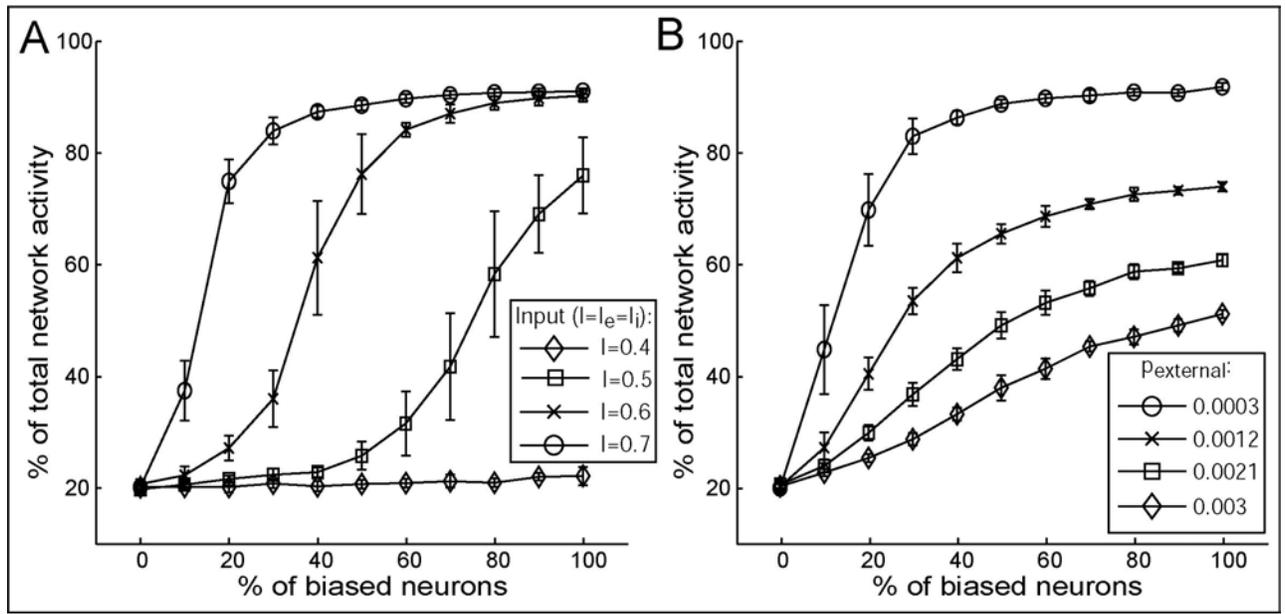

**FIGURE 5**